\documentclass{article}
\usepackage{spconf,amsmath,graphicx}
\usepackage{times}
\usepackage{amssymb}
\usepackage{textcomp}
\usepackage{verbatim}
\usepackage{subcaption}

\title{A Multiple Decoder CNN for Inverse Consistent 3D Image Registration}

\name{Abdullah Nazib$^1$, Clinton Fookes$^1$, Olivier Salvado$^2$,
	Dimitri Perrin$^1$}
\address{$^1$Queensland University of Technology, Brisbane, Australia\\
	$^2$CSIRO, Data61\\
	{\tt\small dimitri.perrin@qut.edu.au}}
\begin{document}
%\ninept
%
\maketitle
\begin{abstract}
The recent application of deep learning technologies in medical image registration has exponentially decreased the registration time and gradually increased registration accuracy when compared to their traditional counterparts. Most of the learning-based registration approaches considers this task as a one directional problem. As a result, only correspondence from the moving image to the target image is considered. However, in some medical procedures bidirectional registration is required to be performed. Unlike other learning-based registration, we propose a registration framework with inverse consistency. The proposed method simultaneously learns forward transformation and backward transformation in an unsupervised manner. We perform training and testing of the method on the publicly available LPBA40 MRI dataset  and demonstrate strong performance than baseline registration methods. 
\end{abstract}
\section{Introduction}
\label{sec:intro}
Image registration is a fundamental step in analyzing medical images. Many medical image analysis procedures require local or global alignment of the images into a common space to understand the progression of diseases, changes in tissue structure etc. Traditional methods solve the alignment problem as an optimization problem. The iterative procedure of solving the optimization problem is often caught in local minimum and is unable to provide optimum alignment. In the case of deformable registration, carefully designed image comparison metrics are required. The application of deep learning in many medical image analysis task significantly improves the performance and accuracy which were difficult to achieve previously. 
Due to the challenge of generating ground-truth data, most deep learning-based image registration methods developed recently are unsupervised methods \cite{Balakrishnan2018}. Several proposed registration methods \cite{Balakrishnan2018,Rohe} consist of a convolutional auto-encoder \cite{Ronneberger2015} with skip connections to generate a 3D deformation field and use the Spatial Transformer network to warp the source image towards the target image. The loss is then calculated between the warped source image and the target image. During training these methods learn forward deformation field by randomly selecting the source image and a fixed atlas image. The problem therefore are twofold: firstly, the learned deformation/transformation is only one directional. And this approach is inefficient when backward transformation is required; Secondly, these methods are only applicable for population studies where scanned images are only be registered with a standard atlas image. Any changes to the atlas image require further training. In contrast to these approaches, here we propose a novel multiple-decoder architecture that learns forward and backward transformations simultaneously in each training iteration. 

\section{Related Works}
Deep learning-based image registration methods can be divided into two different approaches. In the first approach, deep networks are used to estimate the similarity between moving and fixed images \cite{Cheng2016,Simonovsky2016} in an optimization framework. 
In second approach, deep-learning frameworks are used directly to estimate registration parameters instead of just similarity measure. Methods in this category include \cite{Sokooti,Rohe}.
Balakishnan et al. proposed VoxelMorph (VM), a fully unsupervised registration method and achieved state-of-the art performance in 8 different datasets \cite{Balakrishnan2018}. A U-net FCN architecture is combined with a spatial transformation network (STN) for interpolation. The network  is optimized by cross-correlation loss between the fixed and interpolated moving image. Rigorous training and testing on eight different MRI datasets with around eight thousand 3D brain images make this method a viable competitor of conventional state-of-the-art. In this paper, VoxelMorph is one of the baseline methods used for comparison.

\section{Method}
In this paper we propose a multiple decoder  U-net architecture that learns to generate both forward and backward deformation by two decoders branches. To learn both deformation fields, the network is trained by reciprocal cross-correlation losses and cyclic loss. Gradual down-sampling of the last layer in the encoder adversely affects the deformation generation. We hypothesize that an inverse deformation computation will help the network to learn a more accurate forward deformation field. The reciprocal similarity loss plays a pivotal role in this operation, as it ensures that the network generates more accurate deformation filed. Thus the proposed inverse consistent U-net not only provides better registration accuracy but also ensures reversibility between the estimated deformation fields.

\subsection{Inverse Consistent Network Architecture}
\begin{figure}
	\centering
	\includegraphics[clip,width=\linewidth]{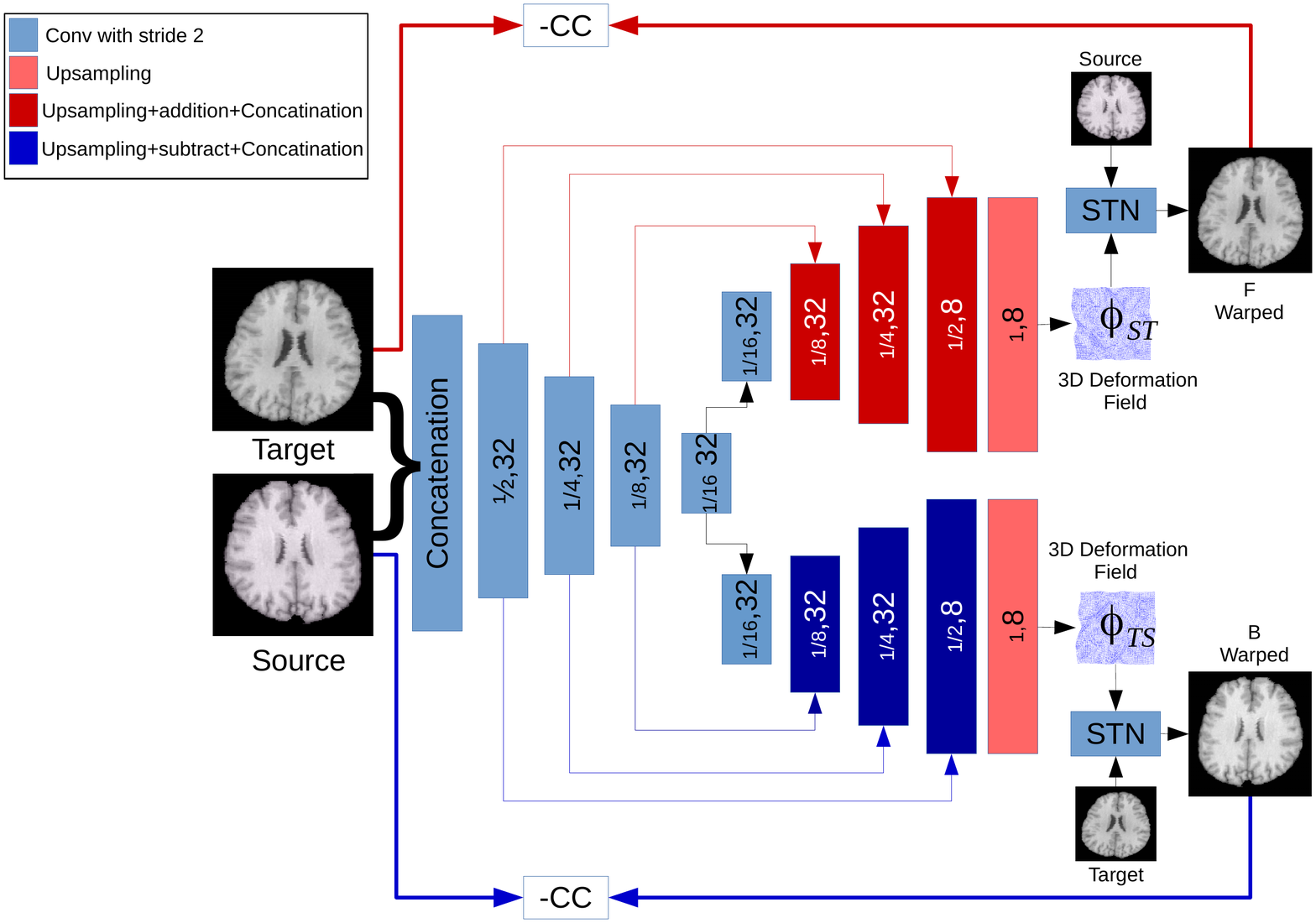}
	\caption{Proposed inverse consistent network architecture with two decoders. 
	Each decoder consists of an upsampling layer, 3 forward/backward computation block and 3 more convolution layers.}
	\label{ch5:fig:inverse_net}
\end{figure}

The architecture of our inverse consistent network is shown in Figure \ref{ch5:fig:inverse_net}. The network consists of an encoder like VM U-net, and two decoder \cite{Ronneberger2015}. In our experiment, the input to the encoder is of size $192\times224\times192\times2$ by concatenating source and target images. The encoder branch has four layers, each of which is a convolution layer of $3\times3\times3$ kernel with stride 2 and 32 output channel. 
Among the two decoder branches, one branch is responsible for forward flow computation and the other one is for backward flow computation. The forward-decoder has one simple convolution layer, three forward computation blocks (red blocks in Figure \ref{ch5:fig:inverse_net}) with 32,32 and 8 channels, and two additional convolution layers with 8 and 3 channels.   
Each forward computation block contains an upsampling layer, an addition layer and concatenation layer. The addition layers adds the same sized features from the encoder branch and then concatenates them. The architecture of the backward-decoder is exactly the same as the forward-decoder with the exception of subtraction layers in the backward computation blocks (blue blocks in Figure \ref{ch5:fig:inverse_net}).  

\subsection{Inverse Consistent Loss}
Since the inverse consistent network has two decoder branches responsible for forward and backward flows, the network is to be trained by two similarity losses. The forward-decoder generates the forward flow which is then used to warp the source image. The similarity is computed between the target image and warped-source image by cross-correlation. In the backward decoder branch, the generated backward flow is used to warp the target image and then cross-correlation (CC) is calculated between the warped-target image and the input source image. 
The final similarity loss consists of negative cross-correlations calculated for the two decoder branches. The forward CC tries to maximize the similarity between the warped source image and the target image while the backward CC tries to maximize the similarity between the warped target and source. The overall similarity loss is then expressed as,
\begin{equation}
\mathcal{L}_{similarity} = -CC(S \circ \phi_{ST}, T)-CC(T \circ \phi_{TS},S).
\end{equation}
Since the two decoding branches of the network generates two flow fields simultaneously, a cyclic loss (from \cite{CycleGAN2017}) is also added to learn inverse consistency,

\begin{equation}
\mathcal{L}_{cycle} =||((T \circ \phi_{TS}) \circ \phi_{ST}) - T||_{1} + ||((S \circ \phi_{ST})\circ \phi_{TS}) - S||_{1}.  
\end{equation}

Then the overall training loss, 
\begin{equation}
\mathcal{L}_{Loss} = \mathcal{L}_{similarity}+ \mathcal{L}_{cycle}.
\end{equation}

\section{Experiments}
\subsection{DataSet}
To evaluate the applicability and performance of the proposed network, the publicly available LPBA40 dataset \cite{Shattuck2008} is used to train and test the network. The dataset contains T1-weighted MR images of 40 healthy young adults. For each subject, the dataset contains an MR image of the brain with skull, one image without skull (skull stripped) and a binary mask of the brain.
To make the dataset consistent and affine aligned to a known space, we select subject 01 of the dataset as the standard space and affine aligned all 39 skull-stripped brain images to it using the ANTS tool \cite{Avants2008}. To make the network able to capture any arbitrary deformation, we follow the pairwise training formulation  $N\times(N-1)$. For the testing, we select 10 brains as the test data. For 10 test brains we generate 90 brain pairs and align all of them with the proposed method and two baseline methods. 

\subsection{Evaluation Metrics}
The proposed architecture is compared with its architecturally closest counterpart VoxelMorph (VM) and with the iterative method ANTS or Symmetric Normalization Tool (SyN). Both proposed method and VM are trained and tested on the same dataset LPBA40. The performance of all methods is measured by Dice similarity score (DSC). DSC values measure the overlap correspondences between the target and the registered image.
Apart from the dice scores, we also calculated the Binary intensity ratio (BIR) as,
\begin{equation}
BIR  = 1 - \frac{IC_{Binary Difference Label}}{IC_{Binary target Label}} 
\label{Eq:BIR}
\end{equation}
where IC = Non-zero intensity count.
To calculate the BIR score from all 90 pairs, we calculated the difference between segmentation label of the target image and the warped source segmentation label, and then binarized the difference label and target label to use them in Eqn \ref{Eq:BIR}. The BIR score represents the transformation capability of the registration tool. A higher BIR score represents better deformation and registration.
The registration time is also an indicative factor of the registration algorithms. We also compare the time taken by each method to register a pair of images.

\subsection{Implementation}
The network is developed in Keras with a tensorflow backend and is trained  and  tested in a High Performance Computing  environment with 64GB  RAM, 12GB  Video RAM in Tesla K40m GPU and a single core 2.66GHz 64bit Intel Xeon processor. We train the network network by ADAM optimizer with a learning rate 1e-4.  

\section{Results}
\subsection{Registration Accuracy}
\begin{figure*}[h]
	\begin{center}
		\begin{minipage}[t]{16cm}\includegraphics[width=\textwidth]{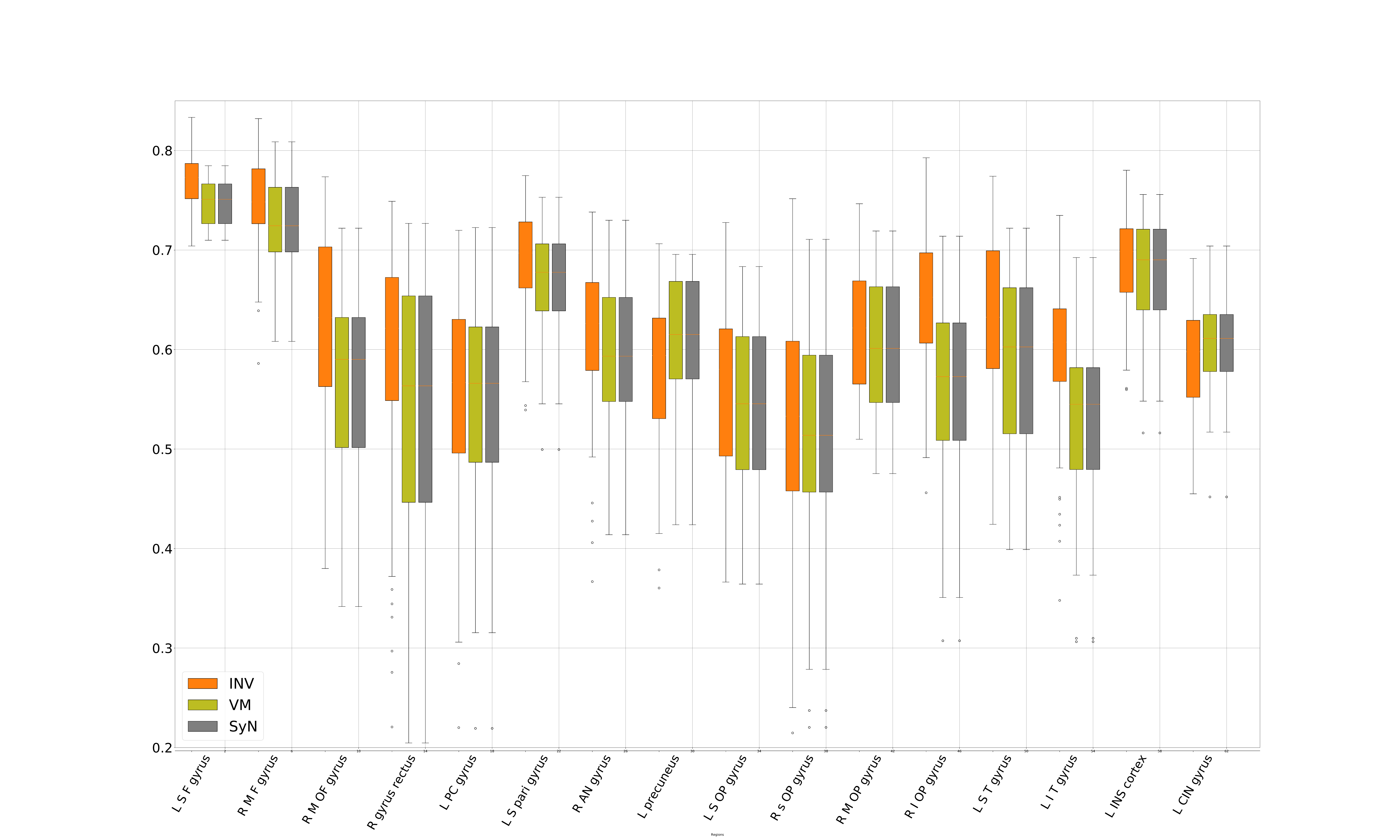}\end{minipage}
		%\begin{minipage}[t]{16cm}\includegraphics[width=\textwidth]{images/Boxplot1.pdf}\end{minipage}
		\caption{Boxplots of Dice Scores of Different Anatomical Structures for InverseNet, Voxelmorph and ANTS}
		\label{Fig:Dice_scores}
	\end{center}
\end{figure*}

The overall performance of three registration tools are shown in Figure \ref{Fig:Overlap} and Figure \ref{Fig:Dice_scores}. In LPBA40 dataset, 56 brain regions are manually labelled. For Dice score estimation, we compare the performance of three methods on all 56 brain regions. In Figure \ref{Fig:Dice_scores}, the boxplots show the distribution of dice scores performed by each registration tool. The first x-axis represents the acronym of the names of the registration methods being compared. The red marked acronyms are representing the best performing tool. The brain region names are labelled in the second x-axis. 
From Figure \ref{Fig:Dice_scores}, its clear that the proposed method obtained best Dice scores in almost all brain regions. The ANTS outperformed the proposed method only in 5 regions. Voxelmorph on the other hand remains the lowest performer among the three. 

In Table \ref{Tab:Accuracy} the mean and standard deviation of BIR scores of 90 brain pairs are presented. The Table \ref{Tab:Accuracy} shows the highest mean BIR score and lowest standard deviation obtained by the proposed InverseNet method while ANTS scored second and VoxelMorph is the lowest performer.     

%%%%%%%%%%%%%%%%% Statistics Table %%%%%%%%%%%%%%%%%%%%%
\begin{table}[t]
	\begin{center}
		\caption{\small  Registration Statistics}
		\label{Tab:quantitative}
		\small
		\begin{tabular}{l c c}
			\hline
			Methods &Mean BIR &BIR Std \\
			\hline
			ANTS				    &0.493835919    &0.0716950053\\
			VoxelMorph              &0.4900602782   &0.0740086096\\
			InverseNet	    		&\textbf{0.5576287543}   &\textbf{0.0701566959}\\
			\hline
			\label{Tab:Accuracy}
		\end{tabular}
	\end{center}
	
\end{table}
%%%%%%%%%%%%%%%%%%%% Images %%%%%%%%%%%%%%%%%%%%%%%%%%
\subsection{Qualitative Comparison}
\begin{figure}[h!]
	\begin{center}
	    \begin{minipage}[t]{3cm}
			\includegraphics[width=2.9cm,height=3.1cm]{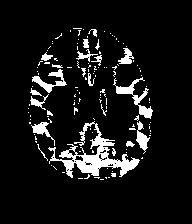}
				\subcaption{INV}
		\end{minipage}%
		\begin{minipage}[t]{3cm}
			\includegraphics[width=2.9cm,height=3.1cm]{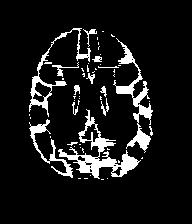}
				\centering\subcaption{VM}
		\end{minipage}%
			\begin{minipage}[t]{3cm}
			\includegraphics[width=2.9cm,height=3.1cm]{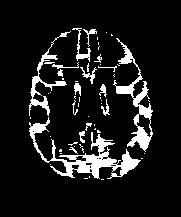}
			\centering\subcaption{SyN}
		\end{minipage}%
			\vfill
		\begin{minipage}[t]{3cm}
			\includegraphics[width=2.9cm,height=3.1cm]{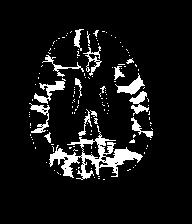}
				\centering\subcaption{INV}
		\end{minipage}%
		\begin{minipage}[t]{3cm}
			\includegraphics[width=2.9cm,height=3.1cm]{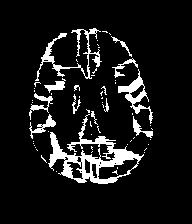}
				\centering\subcaption{VM}
		\end{minipage}%
		\begin{minipage}[t]{3cm}
			\includegraphics[width=2.9cm,height=3.1cm]{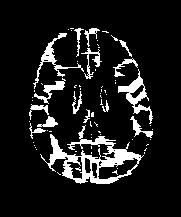}
			\centering\subcaption{SyN}
		\end{minipage}%
		\vfill
		\caption{Visual comparison of registration methods on two random brains: brain1 in 1st Row and brain2 in 2nd Row.}
		\label{Fig:Overlap}	
	\end{center}
\end{figure}

The qualitative performance of the registration tools are shown in Figure \ref{Fig:Overlap}. Figure \ref{Fig:Overlap} is a binarised representation of the difference between the target label and warped label image. The Higher the registration accuracy, the smaller the white non-overlapping region. This figure represents the performance of ANTS, VM and INV on different brains. Among the three methods, the size of non-overlapping (white) regions in ANTS and VM are higher than the INV which indicates superior performance of the proposed INV. The thin boundary lines in the INV registered images compared to other two methods indicates better overlap hence better registration performance in the boundary region of the brains.            

\begin{comment}

\begin{figure*}[h]\ContinuedFloat
	\begin{center}
		\begin{minipage}[t]{16cm}\includegraphics[width=\textwidth]{images/Boxplot2.pdf}\end{minipage}
		\begin{minipage}[t]{16cm}\includegraphics[width=\textwidth]{images/Boxplot3.pdf}\end{minipage}
	\end{center}
\end{figure*}
\begin{figure*}[h]\ContinuedFloat
	\begin{center}
		\begin{minipage}[t]{16cm}\includegraphics[width=\textwidth]{images/Boxplot4.pdf}\end{minipage}
		\begin{minipage}[t]{16cm}\includegraphics[width=\textwidth]{images/Boxplot5.pdf}\end{minipage}
		\caption{Boxplots of Dice Scores of Different Anatomical Structures for InverseNet, Voxelmorph and ANTS}
		\label{Fig:Dice_scores}
	\end{center}
\end{figure*}
\end{comment}

\subsection{Registration Time}
Since ANTS has no GPU version available, ANTS is tested in the same HPC environment without GPU and with four resolution level with 1000 iterations in each level. In this setting, ANTS takes 1862 seconds on average to register a pair of brains. Compared to ANTS, both VM and INV perform registration with a much smaller time of 4 sec and 6 sec respectively. Since the architecture of the VM contains a single decoder, its running time is smaller compared to the proposed INV. Despite its smaller running time, VM is required to re-run to obtain the inverse transformation by swapping the target and source image, while the proposed INV can do both registrations in a single run.          
\begin{comment}

\begin{table}[t]
	\begin{center}
		\caption{\small  Registration Statistics}
		\label{Tab:quantitative}
		\small
		\begin{tabular}{l c}
			\hline
			Methods &Registration Time \\
			\hline
			ANTS				    &1862 sec\\
			VoxelMorph              &4 sec \\
			InverseNet	    		&6 sec \\
			\hline
			\label{Tab:Time}
		\end{tabular}
	\end{center}
\end{table}
\end{comment}

\section{Conclusion}
In this paper we proposed a deep learning-based inverse consistent image registration method. The proposed method is tested on publicly available MR data and achieved better performance than an existing deep learning-based method and a state-of-the-art iterative method. The ability to register in both directions within a very short time with high accuracy makes it a definite alternative to traditional iterative tools (like ANTS) and deep learning-based VoxelMorph method. 
Despite its high accuracy and efficiency, the proposed method can only perform local alignment and requires rigid alignment to be performed before the local alignment using iterative registration tools. We hope our future research direction will alleviate this dependency.
\bibliographystyle{IEEEbib}
\bibliography{refs}
\end{document}